\documentclass[preprint,12pt]{elsarticle}
\usepackage{amssymb}
\begin{document}

\begin{frontmatter}

\title{ Deviation From $\Lambda$CDM With Cosmic Strings Networks}


\author[ctp]{Sumit Kumar}
\ead{sumit@ctp-jamia.res.in}
\author[hri]{Akhilesh Nautiyal}
\ead{akhilesh@hri.res.in}
\author[ctp]{Anjan A Sen}
\ead{aasen@jmi.ac.in}

\address[ctp]{Centre For Theoretical Physics, Jamia Millia Islamia, New Delhi, 110025, India}
\address[hri]{Harish-Chandra Research Institute, Jhunsi, Allahabad, India}

\begin{abstract}

In this work, we consider a network of cosmic strings to explain possible  deviation from $\Lambda$CDM behaviour. We use different observational data to constrain the model and show that a small but non zero contribution from the string network is allowed by the observational data which can result in a reasonable departure from $\Lambda$CDM evolution. But by calculating the Bayesian Evidence, we show that the present data still strongly favour the concordance $\Lambda$CDM model irrespective of the choice of the prior. 

\end{abstract}


\end{frontmatter}


\section{Introduction}
One of the greatest mysteries in cosmology today is the nature of dark energy in the Universe. According to the concordance cosmology,  this makes up around $70\%$ of the total energy density of the universe and causes the universe to accelerate around the present epoch \cite{review}. Yet, there is no strong theoretical as well as observational indication to pin point the nature of this dark component. In fact we still do not  know whether the mysterious late time acceleration is due to the presence of dark energy or due to any modification of gravitational laws on large scales.

Nevertheless, majority of the observational data \cite{obs, suzuki, komatsu, Eisenstein} support the concordance $\Lambda$CDM model as a possible explanation for the late time acceleration.  In this model, the presence of the cosmological constant ($\Lambda$)  poses two serious problems. First one is related to the fine tuning of the value of $\Lambda$ to a ridiculously small number ($\sim 10^{-120} M_{pl}^{4}$ in natural units) necessary for the present day cosmic acceleration. The other one is related to its dominance over the matter component of the universe precisely at the present epoch. Any small departure from these two conditions will result in a cosmological evolution that is different from the observable Universe. These conceptual problems dilute the superiority, the $\Lambda$CDM model enjoys over other dark energy models as far as observational results are concerned.

Interestingly, current observational data also allow deviation from the $\Lambda$CDM behaviour. Dark energy models with $w_{de} = \frac{p_{de}}{\rho_{de}} \neq -1$ are also allowed by current observations \cite{obs, suzuki, komatsu, Eisenstein}. Although the allowed deviation  from $w_{de} = -1$ is not large, but it is still detectable at the precision levels of current and future observational set ups. Hence efforts are on to construct models that can explain this deviation. Almost all the approaches for this purpose, assume that $\Lambda$ is exactly zero in our universe, and an evolving  dark energy is solely responsible for the late time acceleration of the universe. This includes possibilities likes quintessence \cite{quint}, k-essence \cite{kess}, an arbitrary barotropic fluid like Chaplygin gas or its various generalization and many more \cite{gcg}. All these models are generally phenomenological without any strong theoretical motivations and they also burden us with the challenge of explaining several other model parameters.
 
 The other option is to assume the existence of a small but non-zero $\Lambda$ as in $\Lambda$CDM model but also to consider some extra component to explain the observed deviation from $\Lambda$CDM behaviour \cite{aas}. This may be motivated by the fact that string-landscape model may explain the existence of a small but non zero $\Lambda$ in our present universe \cite{SL}.  But it is important to have well motivated candidate that can explain the deviation from $\Lambda$CDM; otherwise this approach will have the same shortcomings which are  present in standard dark energy models mentioned above.
 
In this study, we propose one such model. There are models where dark energy is assumed to be a perfect fluid having equation of state $w = \frac{p}{\rho} <0$. But one major problem for such models is due to its sound velocity $c_{s}^2 = w < 0$ which causes instabilities on the small scales. It was then proposed to include rigidity in the fluid to make it an elastic solid \cite{spergel}. In this case, if the rigidity is sufficiently large, $c_{s}^{2} > 0$ even if $w < 0$. 
 
\section{$\Lambda$CDM plus Cosmic Strings}
 
A solid perfect fluid can arise in different theories. One such possibility is the network of cosmic strings \cite{dp, pm, av}. These are topological defects that are predicted to be formed during the phase transitions in the early universe. Cosmologically they are interesting because if such string configurations are allowed in any particle physics models, they will eventually form string network and may survive till today. Hence detecting these cosmic strings can give us important clues about physics at very high energy scales. If they exist, they can have effects on CMB anisotropies \cite{cmb}, early reionizations \cite{reion}, gravitational waves \cite{gw} as well as in cosmological 21 cm observations \cite{21}. One can put observational bound on the gravitational interaction of these strings through dimensionless quantity $G\mu$ where $G$ is Newton's constant and $\mu$ is the tension of the string. The latest bound on this parameter as provided by the Planck measurement \cite{planck} is $G\mu < 1.5 \times 10^{-7}$.

If these strings are produced during the phase transition at the electroweak scale, they can be very light. But there can be many such strings and their combined energy density can be finite. One interesting feature in the evolution of string network is the intercommutation. Through this process long strings can be cut into smaller strings as well as in loops which evaporate into gravitational waves. At the end of this intercommutation, the string network approaches a scaling regime where the string energy density exactly scales as $a^{-3}$ in the matter dominated era or $a^{-4}$ in the radiation dominated era. This has been shown in the numerical simulations \cite{simul}. In such a scenario, strings can never accelerate the universe. 

On contrary, if the strings do not intercommute or pass through each other, then the network can be frozen in comoving coordinates. These are called fraustrated network of strings. Such network of strings will have the equation of state $w=-1/3$ and hence their energy density will go as $\rho_{s} \sim a^{-2}$. A phase transition around few TeV and a string separation today of the order of few A.U. will result $\Omega_{s}$ today in the range of 0 and 1\cite{dp}.  So a Universe containing fraustrated network of cosmic strings may result in a possible deviation from $\Lambda$CDM model. We shall show in our subsequent study that a small but non zero contribution from this string network can be sufficient for this purpose.  We should also mention that although the existence of these strings networks in cosmological scales are still not confirmed, they can be seen in the condensed matter systems like  biaxial nematic liquid crystals \cite{crystal}. 

Obviously the question is whether in any reasonable particle physics model, we get a fraustrated network of strings. Usually the equation of state of a cosmic string network is related to its root-mean squared (rms) velocity as \cite{KT}
 
 \begin{equation}
 w_{s} = \frac{1}{3} (2 v_{s}^{2} -1).
 \end{equation}
As we have discussed in the previous paragraphs, for a fraustrated network of cosmic strings $w_{s}=-1/3$.  This can be achieved under the limit $v_{s}\rightarrow 0$. It was shown by several numerical simulations that abelian string networks intercommute and give rise to scaling solution without fraustration \cite{vachas}. But later on it was shown that in non-abelian string network, intercommutation may be prevented due to topological constraints and stable fraustrated network of strings may form \cite{dp}. More recent numerical simulations show that in the string dominated universe $v_{s}^{2} \leq 0.17$ \cite{martins,avelino} and can never be exactly zero for nonintercommutating strings.  In this case, the string network can never be conformally stretched to the Hubble scale preventing them to fraustrate. This can only be achieved if one includes some additional fraustration mechanisms to keep the velocity small. This has been discussed by Sousa and Avelino \cite{sousa}. Presence of a massive junction  \cite{junction}  or a conserved charge \cite{charge} can result such fraustration. But for this the energy density of background matter (radiation or dust) should also dominate over the cosmic string energy density ( that includes the contribution from the fraustration mechanism) and hence this string network alone can never accelerate the Universe \cite{avelino}.

Henceforth, we shall abandon the idea of fraustration and consider scenarios that are more similar to the numerical simulations on string networks ( we should rely on this simulations till we detect cosmic strings). In other words, we shall assume the form for $w_{s}$ as described by equation (1) and also consider non zero $v_{s}$ with $v_{s}^2 \sim 0.17$ \cite{martins}. We should stress that in the absence of any actual observational knowledge for string network, this may be a reasonable assumption. 

In this scenario, we want to see whether a small contribution from such a cosmic string network is allowed by the observational data, that can result observable deviation from the concordance $\Lambda$CDM model.

There is also another simple way that can give a string-like term ($\rho_{s} \sim a^{-2}$) in the Hubble equation.
 A fluid with a constant negative pressure ( $p = -p_{0}$, with $p_{0}$ being a positive constant) behaves like a $\Lambda$CDM model. This can be easily seen by putting the constant $p$ in the energy conservation equation $\dot{\rho} +3 H (\rho +p) = 0$ and then integrate it out to obtain the expression for $\rho$. Let us write the small deviation from the $\Lambda$CDM behaviour around present day ($z=0$) by Taylor expanding  around this constant $p$ behaviour:
 
 \begin{equation}
 p = - p_{0}  + p_{1}z + p_{2} z^2,
 \end{equation}
 
 \noindent
 where we keep terms upto second order in $z$. $p_{1}$ and $p_{2}$ are related to $\frac{dp}{dz}|_{z=0}$ and $\frac {d^{2}p}{dz^{2}}|_{z=0}$ respectively. With this one can integrate the energy conservation equation to get the expression for energy density and pressure as a function of scale factor:
 \begin{eqnarray}
 \rho(a) &=& (p_{0} + p_{1} - p_{2}) + \frac{c}{a^{3}} - \frac{3}{2}\frac{p_{1} - 2 p_{2}}{a} - \frac{3 p_{2}}{a^2}\nonumber\\
 p(a) &=& -(p_{0} + p_{1} - p_{2}) + \frac{p_{1} - 2 p_{2}}{a} + \frac{p_{2}}{a^2}. 
 \end{eqnarray}
 
 \noindent
 Here $c$ is an arbitrary integration constant. For the choice $p_{1}=2 p_{2}$ and $p_{2} < 0$, these gives a mixture of fluids containing $\Lambda$, matter and a fraustrated network of cosmic strings. Obviously this particular choice of relation between $p_{1}$ and $p_{2}$ has no underlying theoretical reason, but depicts a simple way of attaining a string-like equation of state in the FRW Universe.
 
 Hereafter we consider a mixture of fluids consisting a dust like matter (with $p = 0$), a cosmological constant ($p = -  \rho$) and a string network with equation of state given by equation (1) and discuss its various cosmological implications.  We start with the Einstein equations:
 
 \begin{eqnarray}
 \frac{H^{2}}{H_{0}^2}  &=& \Omega_{m0} a^{-3} + \Omega_{\Lambda} + \Omega_{s0} a^{-3(1+w_{s})}\nonumber\\
 2\dot{H} + 3H^2 &=& -3H_{0}^{2}\left[w_{s}\Omega_{s0}a^{-3(1+w_{s})} - \Omega_{\Lambda}\right]
 \end{eqnarray}
 
 \noindent
 Here subscript "0" means values at present ($z=0$, $a_{0} =1$), $w_{s}$ is the equation of state for the network of  strings. Flatness condition demands that $\Omega_{m0} + \Omega_{\Lambda} + \Omega_{s0} = 1$; hence two of these parameters are independent. Henceforth we shall consider $\Omega_{m0}$ and $\Omega_{s0}$. 
 
The equation of state for the combined dark energy fluid ( $\Lambda$+Cosmic String) is now given by
 
 \begin{equation}
 w_{de} = \frac{\Omega_{m0} + \Omega_{s0} -1 +\frac{1}{3}(2v_{s}^2-1)\Omega_{s0} a^{-2(1+v_{s}^2)}}{1-\Omega_{m0} -\Omega_{s0}  +\Omega_{s0} a^{-2(1+v_{s}^2)}}.
 \end{equation}
 
Later, we reconstruct this dark energy equation of state, $w_{de}$, to see how much deviation from $w=-1$ is allowed by the current cosmological observations.

\section{Observational Constraint} 
 
With this dark energy set up,  we use the present observational data to constrain our model. We begin by considering the Type Ia Supernovae  observation which probes directly the cosmological expansion. These observations measure the apparent luminosity of the Supernova explosion as observed by an observer on earth. In cosmological terms, this is given by luminosity distance $d_{L}(z)$ defined  as 

\begin{equation}
d_{L}(z) = (1+z)\int_0^z\frac{dz^{\prime}}{H(z^{\prime})}\end{equation}

\noindent
Using this, we evaluate the distance modulus $\mu$ (which is an observable quantity) as
\begin{equation}
\mu = m-M = 5\log\frac{d_{L}}{Mpc}+25,
\end{equation}
where m and M are the apparent and absolute magnitudes of the Supernovae respectively. We consider the latest Union2.1 data compilation \cite{suzuki} consisting of 580 data points for the observable $\mu$.

Next, we use the observational constraints on Hubble parameter as available in the literature. Recently Moresco et al. \cite{Moresco2012} have compiled the Hubble parameter measurements in the redshift range $0 < z < 1.75$ using the differential evolution of the cosmic chronometers.  They have built a sample of 18 observational data point for $H(z)$ spanning almost $10$ Gyr of cosmic evolution. These values are given in Table 1. To complete the data set, we also use the latest and the most precise measurement of the Hubble constant $H_{0}$ \cite{komatsu}.

\begin{table}[h!]
\begin{center}
\begin{tabular}{llll}
\hline \hline
$z$ & $H(z)$ & $\sigma_{H(z)}$ & Ref.\\
\hline
0.090 & 69 & 12 & \cite{Simon2005}\\
0.170 & 83 & 8 & \cite{Simon2005}\\
0.179 & 75 & 4 & \cite{Moresco2012}\\
0.199 & 75 & 5 & \cite{Moresco2012}\\
0.270 & 77 & 14 & \cite{Simon2005}\\
0.352 & 83 & 14 & \cite{Moresco2012}\\
0.400 & 95 & 17 & \cite{Simon2005}\\
0.480 & 97 & 62 & \cite{Stern2010}\\
0.593 & 104 & 13 & \cite{Moresco2012}\\
0.680 & 92 & 8 & \cite{Moresco2012}\\
0.781 & 105 & 12 & \cite{Moresco2012}\\
0.875 & 125 & 17 & \cite{Moresco2012}\\
0.880 & 90 & 40 & \cite{Stern2010}\\
1.037 & 154 & 20 & \cite{Moresco2012}\\
1.300 & 168 & 17 & \cite{Simon2005}\\
1.430 & 177 & 18 & \cite{Simon2005}\\
1.530 & 140 & 14 & \cite{Simon2005}\\
1.750 & 202 & 40 & \cite{Simon2005}\\
\hline \hline
\end{tabular}
\caption{$H(z)$ measurements (in units [$\mathrm{km\,s^{-1}Mpc^{-1}}$]) and their errors.}
\label{tab:HzBC03}
\end{center}
\end{table}

Next, we use the combined BAO/CMB constraints as derived recently by Giostri et al. \cite{giostri12}.  For this, one starts defining the comoving sound horizon at the decoupling as:

\begin{equation}
r_s(z_*)= \frac{1}{\sqrt{3}} \int_0^{1/(1+z_*)}\frac{da}{a^2 H(a)\sqrt{1+(3\Omega_{b0} / 4 \Omega_{\gamma 0})a} },\label{eq:r_s}
\end{equation}
where $\Omega_{\gamma 0}$ and $\Omega_{b0}$ are the photon and baryon density parameter respectively at present ($z = 0$) and we have assumed $c=1$. Here $z_*$ is the redshift at decoupling and is accurately given by the formula as obtained in \cite{hu96}. We also put $z_* = 1091$ \cite{jarosik11}. The other important quantity is the redshift of the drag epoch ($z_d \approx 1020$), when the photon pressure can no longer support the  gravitational instability of the baryons. 

Next we define the "acoustic scale":

\begin{equation}
l_A=\pi\frac{d_A(z_*)}{r_s(z_*)}\quad,
\end{equation}
where  $d_A(z_*)= \int_{0}^{z_*}dz'/H(z')$ is the comoving angular-diameter distance. One can also define the "dilation scale" as introduced in \cite{Eisenstein}:

\begin{equation}  
D_V(z):=\left[ d_A^2(z)z/H(z) \right]^{1/3}.
\end{equation}

The 6dF Galaxy Survey  and more recently, the WiggleZ team \cite{blake11} measured this quantity  at $0.106$, $z=0.44$, $z=0.60$ and $z=0.73$. Percival et al. \cite{Percival} also measured the quantity $\frac{r_{s}(z_{d})}{D_{V}(z)}$ at $z=0.2$ and $z=0.35$. Combining these measurements with the WMAP-7 measurement of $\l_{a}$ \cite{jarosik11}, one can obtain the combined measurement of BAO/CMB for the quantity $\left(\frac{d_{A}(z_{*})}{D_{V}(z_{BAO})}\right)$. This has been given in \cite{giostri12}. From this one can obtain

\begin{equation}
\chi^2_{BAO/CMB}={\bf X^tC^{-1}X},
\label{chi2baocmb}
\end{equation}
where
\begin{equation}
{\bf X}=\left(
          \begin{array}{cccc}
          \displaystyle\frac{d_A(z_*)}{D_V(0.106)} -30.95 \\
            \displaystyle\frac{d_A(z_*)}{D_V(0.2)} -17.55 \\
            \displaystyle\frac{d_A(z_*)}{D_V(0.35)} -10.11 \\
             \displaystyle\frac{d_A(z_*)}{D_V(0.44)} -8.44 \\
             \displaystyle\frac{d_A(z_*)}{D_V(0.6)} -6.69 \\
              \displaystyle\frac{d_A(z_*)}{D_V(0.73)} -5.45 \\
          \end{array}
        \right)
\end{equation}
and
\begin{equation}
{\bf C^{-1}}=\left(
          \begin{array}{cccccc}
          0.48435 & -0.101383 &-0.164945 &-0.0305703 &-0.097874 & -0.106738\\
           -0.101383 & 3.2882 & -2.45497 & -0.0787898 & -0.252254 & -0.2751\\
           -0.164945 & -2.45497 & 9.55916 & -0.128187 & -0.410404 & -0.447574\\
           -0.0305703 & -0.0787898 & -0.128187 & 2.78728 & -2.75632 & 1.16437\\
          -0.097874 & -0.252254 & -0.410404 & -2.75632 & 14.9245 & -7.32441 \\
         -0.106738 & -0.2751 & -0.447574 & 1.16437 & -7.32441 & 14.5022  \\
          \end{array}
        \right)
\end{equation}
We refer to \cite{blake11, giostri12, Percival} for detail derivation. 

\begin{figure}
\begin{center}
\begin{tabular}{|c|c|}
\hline
 & \\
{\includegraphics[width=2.6in,height=2in,angle=0]{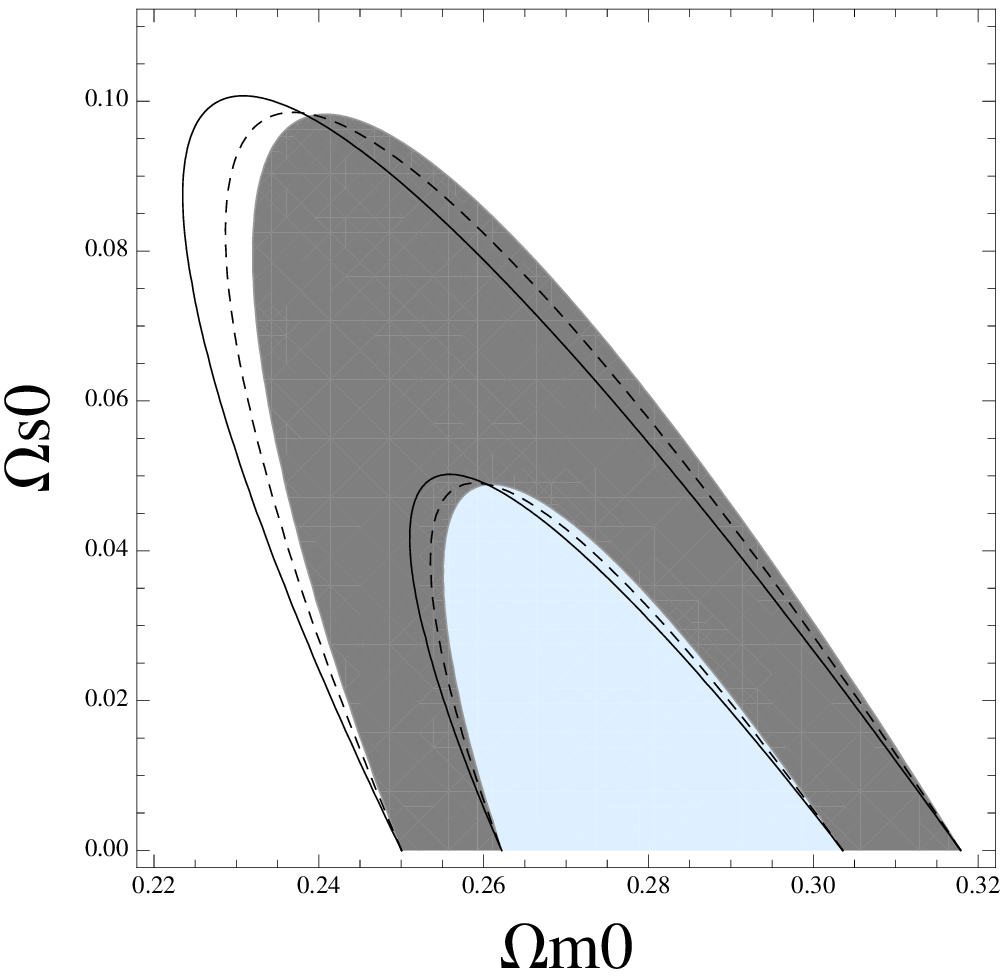}}&
{\includegraphics[width=2.6in,height=2in,angle=0]{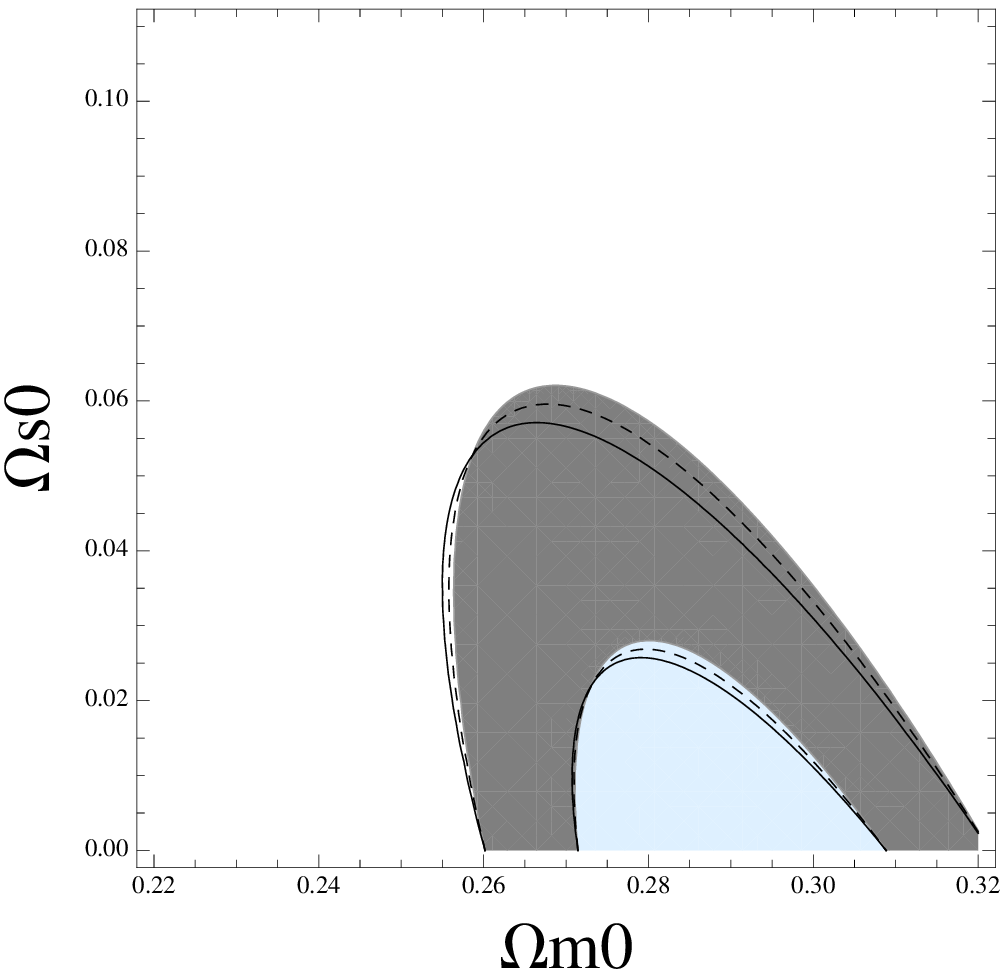}}
\\
\hline

\end{tabular}
\caption{ The $1\sigma$ and $2\sigma$ contour regions in the $\Omega_{m0}-\Omega_{s0}$ plane. The left figure is for flat prior on $\Omega_{m0}$ and the right figure is for Planck prior on $\Omega_{m0}$ ( see text). The shaded contour is for $v_{s} = 0.4$ while the dashed and solid contours are for $v_{s} = 0.45, 0.5$ respectively. In each case the inner contour is for $1\sigma$ and outer contour is for $2\sigma$ confidence level.}
\end{center}
\end{figure}

Recently the Planck Collaboration \cite{planck} has measured the anisotropy in the Cosmic Microwave Background Radiation (CMBR) with unprecedented  accuracy. The parameter $\Omega_{m0}$ has been measured very accurately by Planck and is given by $\Omega_{m0} = 0.315^{0.016}_{-0.018}$ \cite{planck}. We shall use this as a prior ( we call it Planck Prior) in our subsequent calculation. We shall  also use a flat prior for $\Omega_{m0}$ between $0.22$ and $0.38$ and compare the results. For other parameter, $\Omega_{s0}$, we assume a flat prior between  $0$ and $0.4$. As we have mentioned before, we choose a string configuration for which $v_{s}^2 \sim 0.17$. We hence choose three values for $v_{s}$ e.g. $v_{s} = 0.4, 0.45,0.5$ which corresponds to $w_{s} \sim -0.23, -0.2$ and $-0.17$  respectively. 

\begin{figure}
\begin{center}
\begin{tabular}{|c|c|}
\hline
 & \\
{\includegraphics[width=2.6in,height=2in,angle=0]{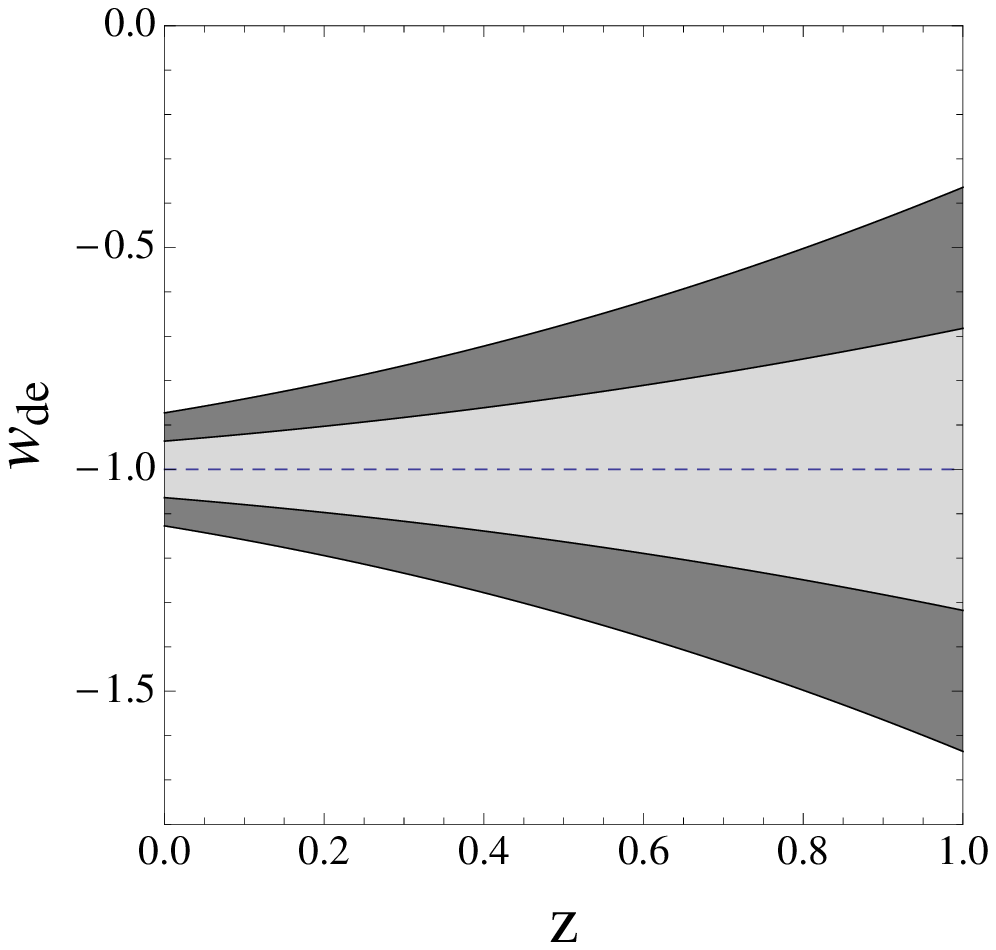}}&
{\includegraphics[width=2.6in,height=2in,angle=0]{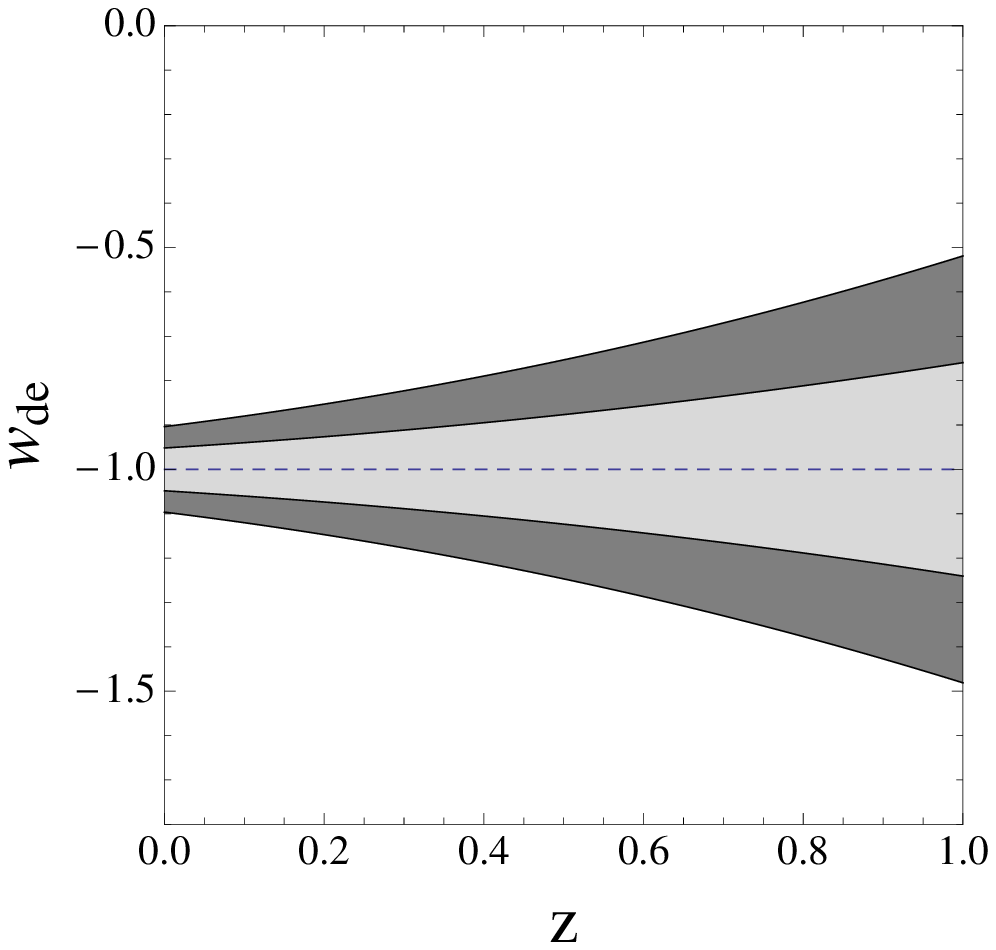}}
\\
\hline

\end{tabular}
\caption{The reconstructed $w_{de}$ behaviour as a function of redshift for $v_{s} = 0.4$. The dashed line is the best fit behaviour. The light grey region is for $1\sigma$ confidence level and the dark grey region is for $2\sigma$ confidence level. The left one is with flat Prior  and the right is for Planck prior on $\Omega_{m0}$ (see text).}
\end{center}
\end{figure}

With these three sets of observational data, we construct the joint likelihood for our model as:

\begin{equation}
-2 log {\cal L} = - 2 Log ({\cal L}_{sn} \times {\cal L}_{bao} \times {\cal L}_{hub} )
\end{equation}.

Subsequently we constrain our model parameters using this likelihood function. We calculate the confidence contours in the $\Omega_{m0}-\Omega_{s0}$ plane for the different values of $v_{s}$ mentioned above both with Planck as well as flat prior for $\Omega_{m0}$. Also knowing the errors for $\Omega_{m0}$ and $\Omega_{s0}$, we use equation (5) to calculate the equation of state for the dark energy $w_{de}$ using the standard technique of error propagation \cite{gregory}. The results are shown in Figures 1.

\begin{figure}
\begin{center}
\begin{tabular}{|c|c|}
\hline
 & \\
{\includegraphics[width=2.6in,height=2in,angle=0]{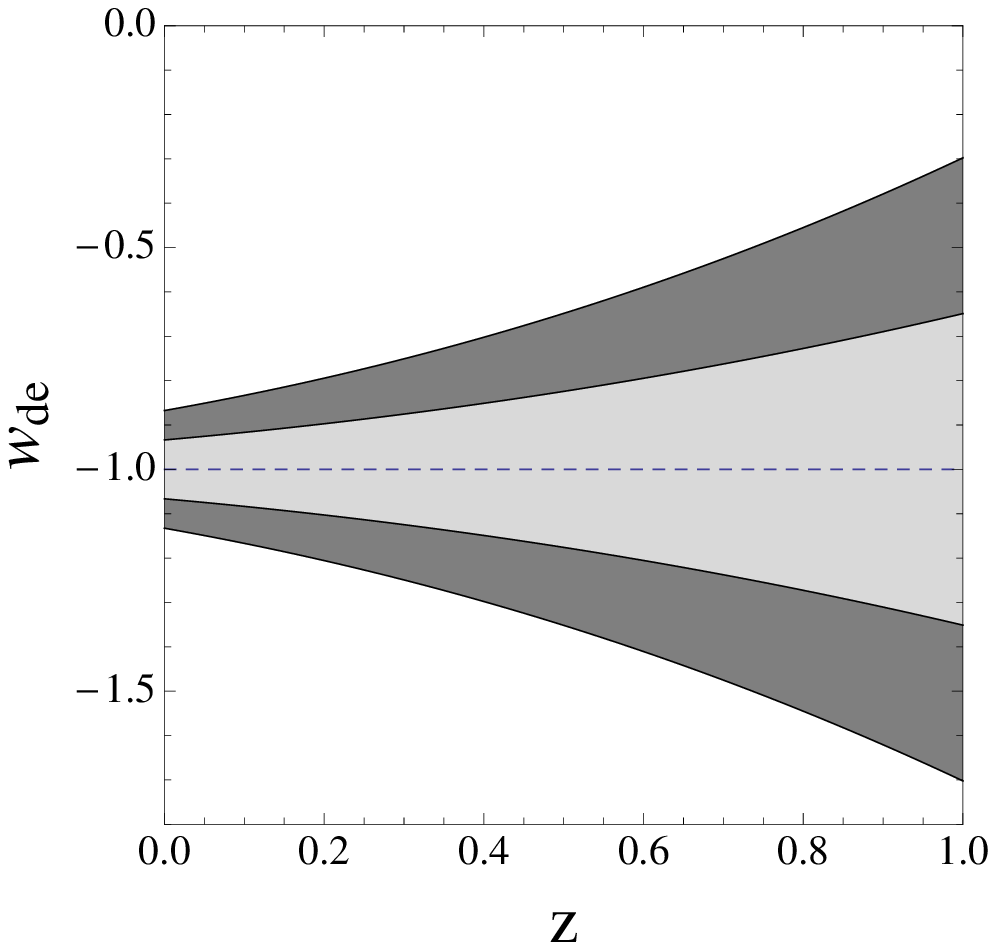}}&
{\includegraphics[width=2.6in,height=2in,angle=0]{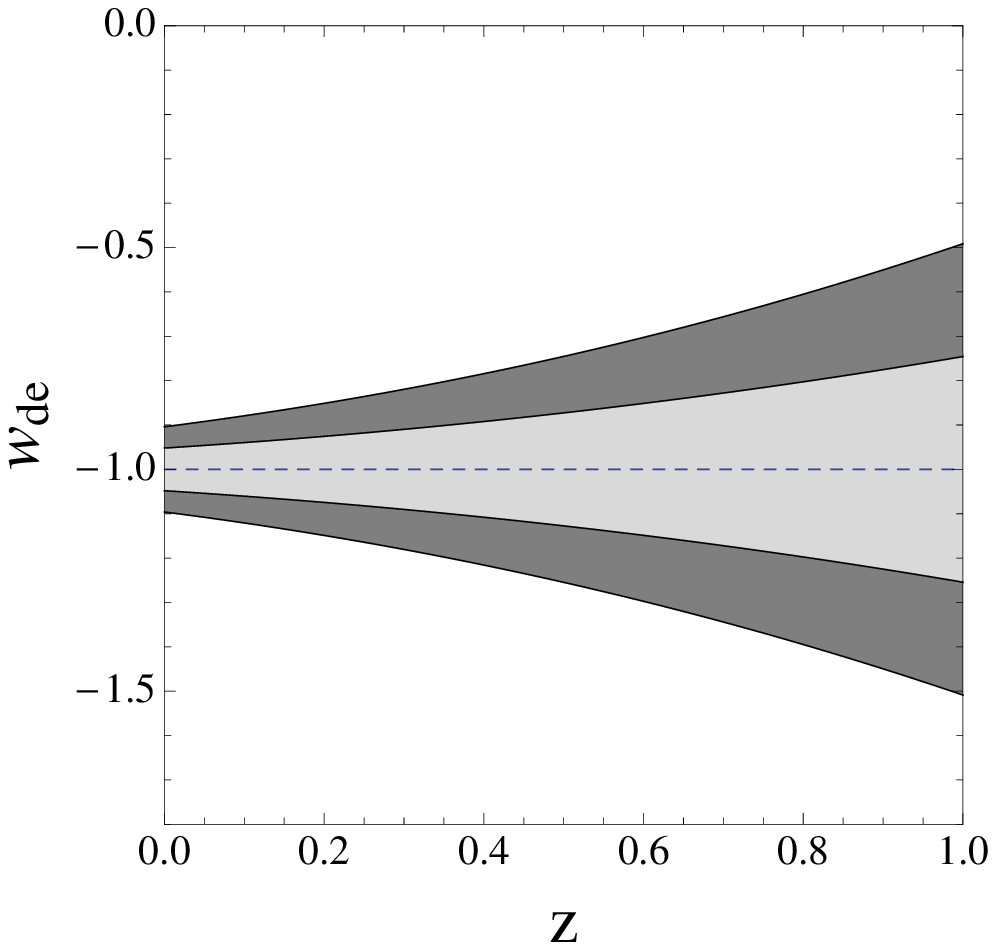}}
\\
\hline

\end{tabular}
\caption{The reconstructed $w_{de}$ behaviour as a function of redshift for $v_{s} = 0.45$. The dashed line is the best fit behaviour. The light grey region is for $1\sigma$ confidence level and the dark grey region is for $2\sigma$ confidence level. The left one is with flat Prior  and the right is for Planck prior on $\Omega_{m0}$ (see text).}
\end{center}
\end{figure}

One can see from this figure that with Planck prior on $\Omega_{m0}$, the allowed value for $\Omega_{s0}$ is smaller than in the case with flat prior on $\Omega_{m0}$. Also the sensitivity on different values for $v_{s}$ is higher for the flat prior case in comparison to the case with Planck prior. For flat prior on $\Omega_{m0}$, the allowed value for string contribution increases with increasing $v_{s}$, whereas for the Planck prior opposite happens although with lesser sensitivity. Also the maximum contribution from the string network can be less than $10$ percent for flat prior on $\Omega_{m0}$ and less than $6$ percent in case of Planck prior on $\Omega_{m0}$.  Hence the contribution from the string network is always very small compared to the matter or dark energy components.

Next we study whether such small contribution from string network can produce appreciable deviation from $\Lambda$CDM model. For that, we reconstruct the effective dark energy equation of state $w_{de}$ given by equation (5). The results are shown in figure 2, figure 3 and figure 4. The results shows that with flat prior on $\Omega_{m0}$, one can have a larger deviation from cosmological constant compared to the case with Planck prior on $\Omega_{m0}$. The variation from cosmological constant increases slightly with increasing values for $v_{s}$ for both Planck prior and flat prior case. For $2\sigma$ confidence limit, the present day value for $w_{de 0}$ can vary between $-1.09$ to $-0.9$ for Planck prior case and between $-1.14$ and $-0.86$ in case of flat prior. It is clear from these results, that even with a tiny contribution from the cosmic string network, there can be reasonable deviation from the cosmological constant that is allowed by the observational data.

To summarise, the current observational data is fully consistent with the presence of a small but non zero contribution from cosmic string networks. This small contribution is enough to produce appreciable change in the behaviour of dark energy equation of state $w_{de}$ at present ($z=0$)  as well as in the earlier times ($z>0$).

With these results, we next study whether our model is preferred over a concordance $\Lambda$CDM model. For this we calculate the Bayesian Evidence which  is defined as \cite{liddle}
\begin{equation}
E = \int {\cal L}(\theta) P(\theta) d\theta,
\end{equation} 
where $\theta$ represents the set of model parameters, ${\cal L}$ represents the Likelihood function and  $P(\theta)$ is the prior probability distribution for parameter $\theta$. According to Jeffrey's interpretation \cite{jef}, if $\Delta \ln E$ between $1$ and $2.5$, it is a significant evidence for a model with higher $E$ while the same between $2.5$ and $5$ is a strong to very strong evidence. If $\Delta ln E$ is more than $5$, the model with higher $E$ is decisively favoured.  

To compare the evidence of our model with $\Lambda$CDM, we calculate the difference in evidence:

\begin{equation}
\Delta \ln E = \ln E_{\Lambda} - \ln E_{cs}
\end{equation}

\begin{figure}
\begin{center}
\begin{tabular}{|c|c|}
\hline
 & \\
{\includegraphics[width=2.6in,height=2in,angle=0]{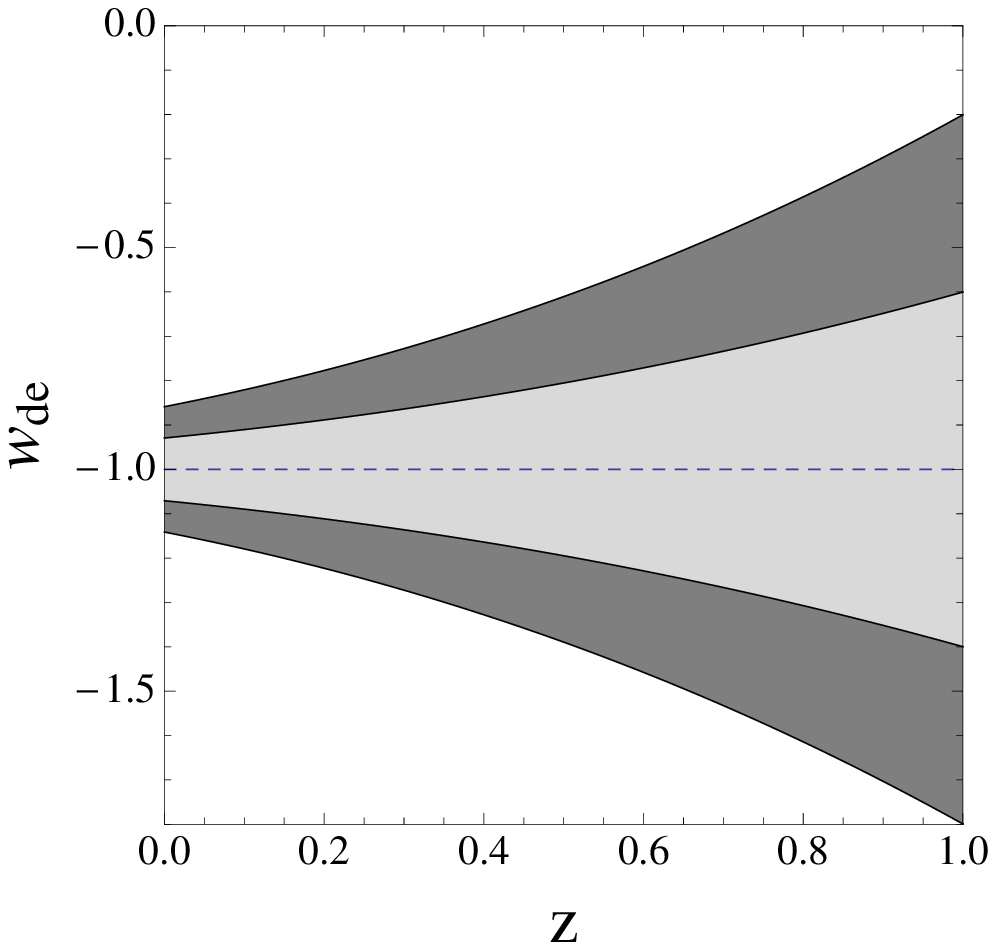}}&
{\includegraphics[width=2.6in,height=2in,angle=0]{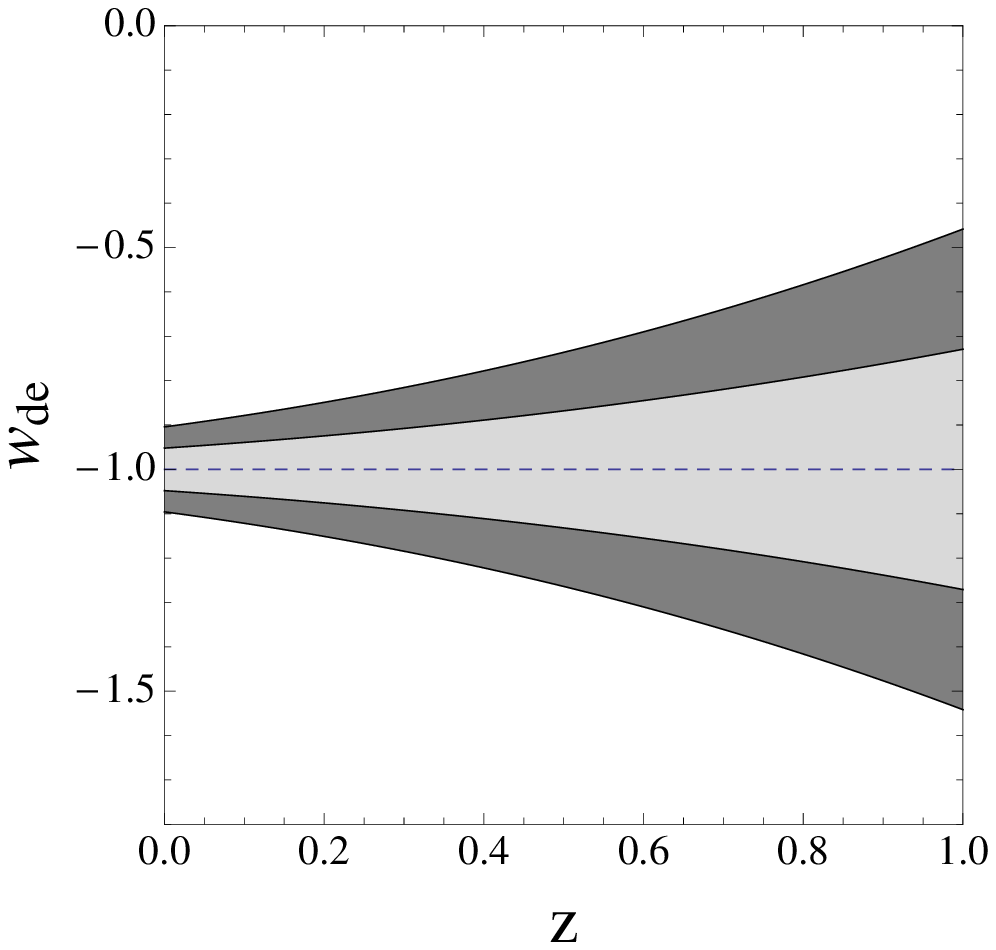}}
\\
\hline

\end{tabular}
\caption{The reconstructed $w_{de}$ behaviour as a function of redshift for $v_{s} = 0.5$. The dashed line is the best fit behaviour. The light grey region is for $1\sigma$ confidence level and the dark grey region is for $2\sigma$ confidence level. The left one is with flat Prior  and the right is for Planck prior on $\Omega_{m0}$ (see text).}
\end{center}
\end{figure}

The results are shown in Table 1. From this one can conclude that although in our model a large deviation from the cosmological constant is allowed by the observational data ( see figures 2,3,4), a concordance $\Lambda$CDM model is still preferred strongly over a $\Lambda$ + cosmic string model by the current observational data.

\renewcommand\arraystretch{1.05}
\begin{table}[t]
\center
\begin{tabular}{|c|c|c|}
\hline
$v_{s}$  &     $ \Delta ln $E (Planck Prior)      &    $ \Delta ln $E (flat Prior)    \\  
\hline
\hline
0.4      &        2.4        & 2.35  \\      
\hline
0.45      &    2.45            & 2.35    \\     
\hline
0.5  &  2.5             &  2.33\\
\hline
\end{tabular}
\caption[evidence]{Difference of Bayesian evidence: $\Delta ln E = ln (E_{\Lambda}) - ln (E_{cs})$}.
\end{table}

 \section{Conclusion}
 
Cosmological observations have now confirmed the late time acceleration of the Universe. The challenge is to explain the source of this acceleration. Although a concordance $\Lambda$CDM model is consistent with different observational results, a small deviation from this $\Lambda$CDM behaviour is also allowed.  Moreover recent results also show that there may be some inconsistencies between different observational results if one assumes $\Lambda$CDM as the correct cosmological model \cite{tension}.  This motivates people to consider scenarios slightly different from an exact $\Lambda$CDM behaviour.

In most of these scenarios, one assumes the $\Lambda$ term to be exactly zero and introduces dynamical scalar fields that can give rise to late time acceleration of the universe. All these models are phenomenological in nature and lacks theoretical understanding ( See \cite{SS} for recent attempt to build a quintessence model in string theory). They not only inherit the problems present in the concordance $\Lambda$CDM model, but also burden us with a number of extra parameters which also have to be fine tuned. We should also stress the fact that except the recently discovered Higgs field in Large Hadron Collider (LHC) \cite{higgs}, no other scalar field has been detected so far that can explain this late time acceleration.
 
This motivates us to look for alternative approaches that may explain the deviation from the $\Lambda$CDM evolution. Formation of cosmic string network is possible during phase transitions in the early Universe. Although, on cosmological scales, we have not detected these strings so far, but they are more commonly observed in condensed matter systems in the laboratory. Keeping this is mind, we propose a cosmological model where a small contribution from cosmic strings network is present in the energy budget of the Universe together with the cosmological constant and other usual matter components. One can consider this network of strings as a perfect gas \cite{KT} and express the overall equation of state $w_{s}$ as a function of its average velocity $v_{s}$. Numerical simulations suggest  that this average velocity can not be exactly zero but is constrained to be $v_{s}^2 \sim 0.17$. 

We show that such a model is completely consistent with current observational data. We put constraint on the model parameters like $\Omega_{s0}$ and $\Omega_{m0}$.  Subsequently calculating Bayesian Evidence, we show that statistically  data still prefer strongly a model with simple cosmological constant and this result does not change much with the choice  of prior. 

To conclude, if the deviation from the $\Lambda$CDM evolution is indeed confirmed, cosmological constant with a small contribution from cosmic string network may explain such deviation. At the theoretical level, we believe this set up has similar appeal as in scalar field models as both can arise in reasonable  particle physics set up though neither has been detected so far. But given the fact that cosmic strings formation has been observed in condensed matter system, and numerical simulations on the dynamics of string networks on cosmological scales have revealed a lot about their properties, if detected, they may be a good alternative to scalar field dark energy models.

\vspace{5mm}
\section{Acknowledgement}

The author SK is funded by the University Grants Commission, Govt.of India through the Junior Research Fellowship. The author AAS acknowledges the funding from SERC, Dept. of Science and Technology, Govt of India through the research project SR/S2/HEP-43/2009. Part of the numerical computations were performed using the Cluster computing facility at the Harish-Chandra Research Institute,Allahabad, India (http://cluster.hri.res.in/index.html).

\end{document}